\documentstyle[aas2pp4]{article}
\tightenlines

\begin{document}

\title{
Asteroseismology and Oblique Pulsator Model of $\beta$ Cephei
}

\author{
H. Shibahashi 
}
\affil{
Department of Astronomy, School of Science,
University of Tokyo,\\
Bunkyo-ku, Tokyo 113-0033, Japan
}
\and
\author{
C. Aerts\footnote{Postdoctoral Fellow of the Fund for Scientific Research,
Flanders, Belgium}
}
\affil{
Instituut voor Sterrenkunde,
Katholieke Universiteit Leuven,\\
Celestijnenlaan 200 B,
B-3001 Leuven, 
Belgium
}

\begin{abstract}
We discuss the oscillation features of $\beta$ Cephei, which is a magnetic 
star in which the magnetic axis seems to be oblique to the rotation axis.
We interpret the observed equi-distant fine structure of the frequency spectrum
as a manifestation of a magnetic perturbation of an eigenmode, which would be a
radial mode in the absence of the magnetic field. Besides these frequency
components, we interpret another peak in the frequency spectrum as an
independent quadrupole mode. By this mode identification, we deduce the mass,
the evolutionary stage, the rotational frequency, the magnetic field strength,
and the geometrical configuration of $\beta$ Cephei.
\end{abstract}

\section{Introduction}  

Beta Cephei (B1III, ${\rm V} = 3.2$, $V_{\rm e} \sin i=25$ km/s) is the
prototype of a group of early-type pulsating stars. 
Since many of the $\beta$ Cephei-type stars show multi-periodic
pulsations, it is expected that a lot of information can be extracted from the
pulsations of these stars. 
Recently, Aerts et al. (1994) and Telting, Aerts, \&
Mathias (1997) investigated the line-profile variability of $\beta$
Cephei in detail, by means of an extensive data set of high-resolution, high
S/N spectra obtained at the Observatoire de Haute Provence during one month,
and deduced that the variations show periodicity with at least five
frequencies. The frequencies obtained by them from the intensity variation in
the Si III $\lambda$ 4574 line are listed in table 1. Note that three
frequencies among them ($f_3$, $f_4$, and $f_5$) are separated from the main
frequency $f_1 = 5.250$ day$^{-1}$ by $-2/6$ day$^{-1}$, $-1/6$ day$^{-1}$, and
$+1/6$ day$^{-1}$, respectively. Hence, we regard that the four peaks (and the
small peak at $f_1 + 2/6$ day$^{-1}$) are a part of a frequency quintuplet with
equal spacing. The central frequency $f_1$ has been identified as the radial
pulsation mode by Aerts et al.\ (1994) and by Telting et al.\ (1997).
The second highest peak in the periodogram, $f_2$ seems to be an independent 
mode, and has been identified as an $\ell=2$ mode by Telting et al. 
(1997).

\begin{table}[hb]
\caption[]{\footnotesize 
Observed frequencies of spectral line variations.
}
\begin{center}
\begin{tabular}{lcc}
\hline\hline
 & Frequencies (day$^{-1}$) & Amplitude \\
\hline
$f_3$ & 4.923 & $6.89\times10^{-4}$ \\
$f_4$ & 5.082 & $1.10\times10^{-3}$ \\
$f_1$ & 5.250 & $1.00$ \\
$f_2$ & 5.380 & $1.41\times10^{-3}$ \\
$f_5$ & 5.417 & $1.40\times10^{-3}$ \\
$f_6$?& 5.583 & $1.62\times10^{-4}$ \\
\hline
\end{tabular}
\end{center}
\end{table}

Besides the pulsations in the luminosity and in the velocity field, a magnetic
field has been reported at several occasions in the past.  For a compilation of
these results, see Veen (1993). No clear periodicity was found in these old
measurements. Very recently however, Henrichs et al. (1999)
re-measured the magnetic field and found a clearly variable field with a period
of 12 days and a semi-amplitude of $90\pm 6\,$G.  The UV wind lines of this
star also reveal a periodic variation with both a 6 day and a 12 day periodic
component (Henrichs et al. 1993), the 6 day period clearly being the dominant
variation.  From these results, it is concluded that the magnetic field of 
$\beta$ Cephei is oblique to the rotation axis.

In this paper,
we try to understand the cause of the quintuplet and to extract
information on the evolutionary stage and the geometrical configuration of
$\beta$ Cephei by using the oscillation data and the most recent magnetic data.
A previous similar study was presented in Shibahashi \& Aerts (1998) but they
still used the older magnetic field measurements.

\section{Oscillations of a Rotating Magnetic Star}

Henrichs et al. (1999) find a variable magnetic field
with a mean value of only $0\pm 5\,$G. We assume here that the dominant
component of the weak magnetic field is dipolar.  Being influenced by such a
magnetic field, the eigenmode, which would be a pure radial mode in the absence
of a magnetic field, is deformed to have an axially symmetric quadrupole
component, whose symmetric axis coincides with the magnetic axis. Hence, the
eigenfunction at the surface is characterized by means of a superposition of
the spherical harmonic with $\ell=m=0$ and that of $\ell=2$ and $m=0$ with
respect to the magnetic axis (Shibahashi 1994, Shibahashi \& Takata 1995);
i.e.,
\begin{equation}
        \left[
        Y_0^{0}(\theta_B,\phi_B)
        + 
        \alpha Y_{2}^{0}(\theta_B,\phi_B)
        \right]
        \exp (i\omega t),
\label{eq:1}
\end{equation}
where $\alpha$ is determined by the magnetic field of the star and the
unperturbed eigenfunction of the mode. Since we are assuming that the magnetic
axis is oblique to the rotation axis, the aspect angle of the pulsation axis
varies with the rotation of the star. Therefore, the contribution of the
quadrupole component of the eigenfunction to the apparent intensity variation
changes with time and produces a quintuplet fine structure with an equal
spacing of the rotation frequency in the power spectrum. Mathematically, a
spherical harmonic expressed in the spherical coordinates with respect to the
magnetic axis $(\theta_B, \phi_B)$ is written in terms of $(2\ell + 1)$
spherical harmonics of the same degree $\ell$ with respect to the spherical
coordinates $(\theta_L, \phi_L)$:
\begin{eqnarray}
        Y_\ell^m(\theta_B, \phi_B)
        &=&
        \sum_{m'=-\ell}^{\ell} \sum_{m''=-\ell}^{\ell}
        (-1)^{m'} d_{0m'}^{(\ell)}(\beta) d_{m'm''}^{(\ell)}(i) \nonumber\\
        & & Y_\ell^{m''}(\theta_L, \phi_L)
        \exp(-im' \Omega t) ,
\label{eq:2}
\end{eqnarray}
where $\{d_{mm'}^{(\ell)}(\beta)\}$ and $\{d_{m'm''}^{(\ell)}(i)\}$ are the
matrices to transform the spherical harmonics expressed in terms of the
spherical coordinates with respect to the magnetic axis, to those expressed in
terms of the spherical coordinates with respect to the line-of-sight. The
explicit form of $\{d_{mm'}^{(\ell)}(\beta)\}$ for $\ell=2$ can be found in
Kurtz et al. (1989). The angles $\beta, i \in [0^\circ,180^\circ]$ are the
angle between the magnetic axis and the rotation axis and that between the
line-of-sight and the rotation axis, respectively.  The observable variation is
obtained by integrating equation (\ref{eq:1}) over the visible disc
($\theta_L=[0, \pi/2]$ and $\phi_L=[0, 2\pi]$) after substituting equation
(\ref{eq:2}) into equation (\ref{eq:1}).  Since the integral with respect to
$\phi_L$ becomes zero unless $m''=0$, the amplitude of the component at $\omega
+ m'\Omega$ is proportional to $d_{0m'}^{(2)}(\beta) d_{m'0}^{(2)}(i)$
(Shibahashi 1986).  If an additional quadrupole component of the magnetic field
is taken into account, the eigenfunction is deformed to have $\ell=1$, 
$\ell=3$, and $\ell=4$
components as well and then the expected power spectrum becomes a nonuplet
(a 9-fold multiplet)
rather than a quintuplet (Shibahashi 1994, Takata \& Shibahashi 1994). 
The additional outer side-components are due to the $\ell=3$ and
$\ell=4$ components, and hence their amplitudes are expected to be much smaller
than the central five components.

\section{Deduction of $\tan\beta$ and $\tan i$}

If the star is rotating with a period of 6 days, 
then the power spectrum of the eigenmode, which would be the radial mode in the
absence of the magnetic field, is expected to reveal a quintuplet with an equal
spacing of 1/6 day$^{-1}$. This can be explained by the observed quintuplet
fine structure of ${f_1}$, $f_3$, $f_4$, $f_5$, and an additional very small
peak in the power spectrum (which we call $f_6$).  The relative ratios of the
side-peaks' amplitudes to the central peak amplitude depend on the strength of
the magnetic field. On the other hand, the relative ratios among the side-peak
amplitudes depend on the angles $\beta$ and $i$. By analyzing them,
we can determine these angles. If we write the amplitude of
the component at $\omega + m'\Omega$ as $A_{m'}$, then
$(A_2+A_{-2})/(A_1+A_{-1})$ is given by (Shibahashi 1986)
\begin{equation}
   {{A_2 + A_{-2}}\over{A_1 + A_{-1}}} 
   = \left|{{d_{02}^{(2)}(\beta)d_{20}^{(2)}(i)}
      \over{d_{01}^{(2)}(\beta)d_{10}^{(2)}(i)}}\right| 
   = {{1}\over{4}}\left|\tan\beta \tan i\right|.
\end{equation}
Substitution of the observed amplitudes into the above equation leads to
$|\tan\beta \tan i| \simeq 2.2$, where we have used $A_2=A_{-2}$ since the
presence of a frequency peak at $f_6$ is not well established.  The factor
$\tan\beta \tan i$ can be independently estimated from the variation in the
magnetic field strength.  Let $r$ be the ratio between the apparent minimum and
the maximum of the field strength,
\begin{equation}
   r \equiv {{B_{\rm obs, min}}\over{B_{\rm obs, max}}}.
\end{equation}
Then it is related with the geometrical configuration by (Stibbs 1950)
\begin{equation}
   \tan\beta \tan i = {{1-r}\over{1+r}}.
\end{equation}
Substitution of the observed values gathered by Henrichs et al. (1999) 
together with their error, 
$r \in [-2.5, -0.6]$, leads to $\tan\beta \tan i \in [-\infty, -2.3]$ or 
$[4.1, \infty]$. This
result is seemingly in contradiction with the one derived from the amplitude
ratio, but we will show in the Sect.\,\ref{ibeta} that the two estimates are
compatible with each other and point to almost the same geometry.

\section{Identification of the Evolutionary Stage}

The second highest peak $f_2$ in the power spectrum has been identified by
Telting et al. (1997) as a mode with $\ell=2$ and $m=+1$. Telting et al. (1997)
have chosen the rotation
axis as the symmetry axis of the pulsation in their work.  If the
magnetic effects on the pulsation dominate over the effects due to the Coriolis
force, as applies here, the symmetry axis of the eigenfunction is the
magnetic axis rather than the rotation axis.  Following Telting et al.'s (1997)
mode identification, we assume that $f_2$ belongs to a quadrupole ($\ell=2$)
mode, but we assume $m=0$ with respect to the magnetic axis.  As will be shown
later, the following conclusion about the evolutionary stage is the same even
in the case of $m\neq 0$.  Under the influence of the magnetic field, this mode
is no longer described by a single spherical harmonic of $\ell=2$ and $m=0$,
and it is deformed to have components of some other $\ell$.  However, the fine
structure of this mode is expected to be difficult to detect, since the
amplitude of the $f_2$ component itself is small.

\begin{figure}[ht]
\epsfxsize=8cm
\centerline{\epsfbox[76 210 792 792]{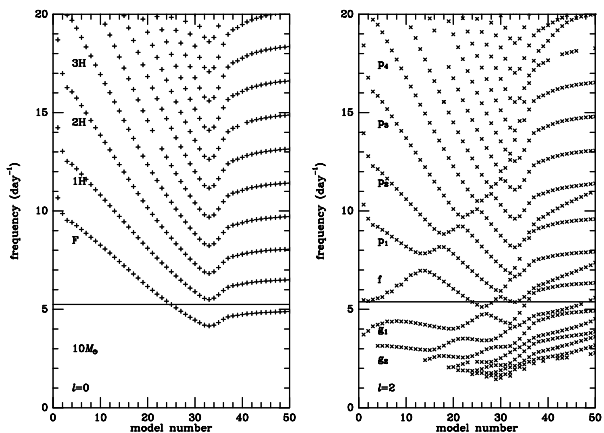}}
\caption{\footnotesize
Frequency variation of radial modes (left panel) and $\ell=2$ modes (right
panel) of a $10 M_\odot$ star with stellar evolution. The vertical scale is in
unit of day$^{-1}$, and the horizontal scale shows the model numbers 
(Model 1 is the
zero-age main-sequence star, and Model 33 is the model at the turning point 
on the HR-diagram.). The solid lines indicate the observed frequencies 
$f_1$ (left panel) and $f_2$ (right panel).
}
\label{fig:1}
\end{figure}

Since we have
two independent frequencies, $f_1$ and $f_2$, we can identify the evolutionary
stage of the star. Figure \ref{fig:1} shows the frequency variation of a $10
M_\odot$ star. The left panel shows the radial mode frequency variation, while
the right panel shows the case of $\ell=2$ modes. As the star evolves from the
zero age main sequency, its radius increases and hence the frequency
decreases. The situation changes during the contraction phase when hydrogen
is exhausted in the stellar center, and the frequency increases again in
the hydrogen shell-burning phase. The frequency variation of the $\ell=2$ mode
is more complicated because the $\mu$-gradient zone around the
convective core grows with evolution as a consequence of conversion of hydrogen
into helium. The frequencies of the modes trapped there become higher. 
When the frequencies of two modes happen to be very close, they never
degenerate but repel each other. This is known as ``avoided crossing'' (cf.
Unno et al. 1989).

\begin{figure}[ht]
\epsfxsize=8cm
\centerline{\epsfbox[76 210 792 792]{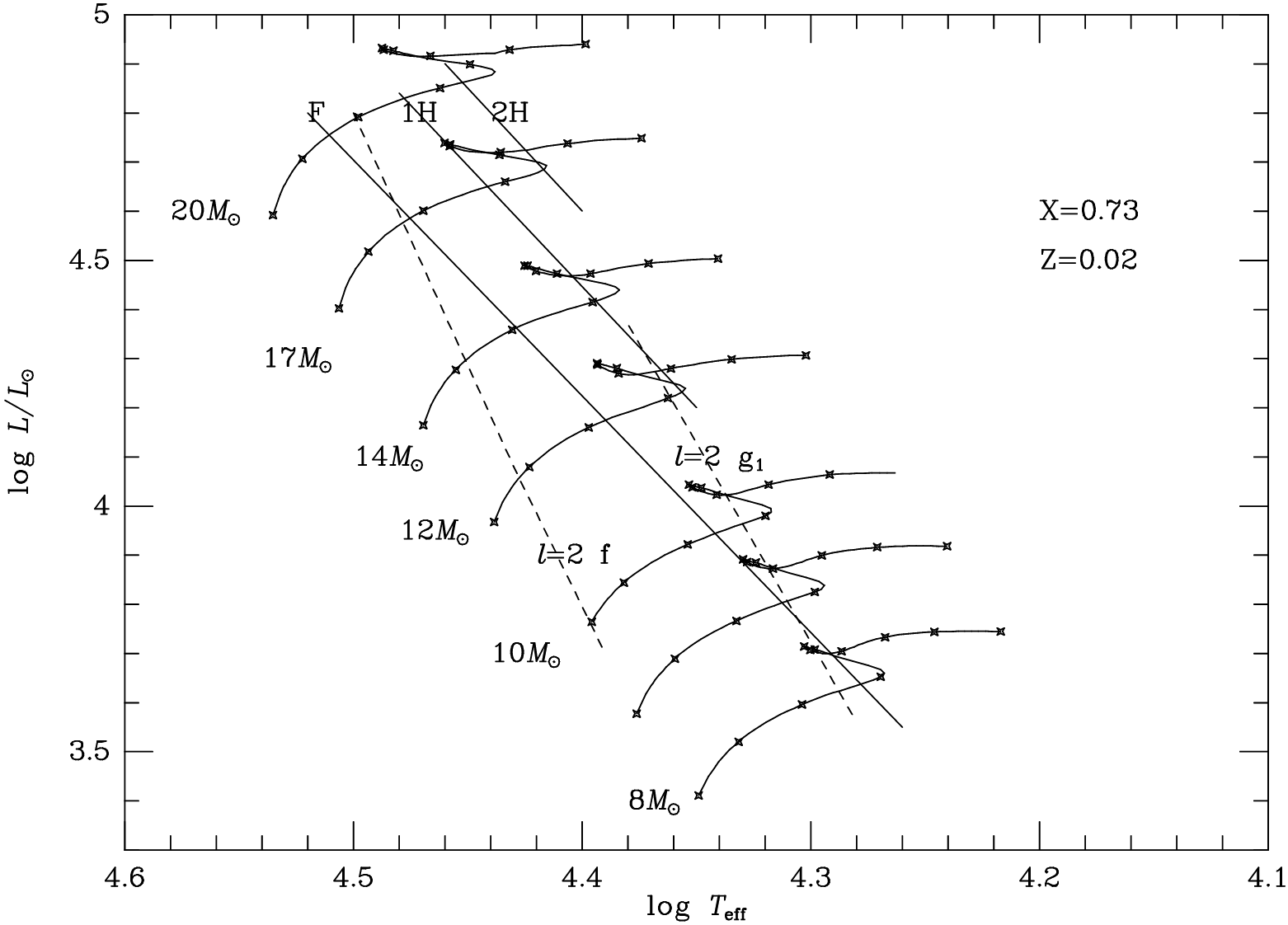}}
\caption{\footnotesize
The HR-diagram of 8--20 $M_\odot$ stars. 
The tick marks on each of the evolutionary tracks indicate Models 1, 10, 20, 
$\cdots$.
The stars whose radial mode frequency
coincides with the observed frequency $f_1$ are connected with the
solid lines, and those whose quadrupole mode frequency matches with
$f_2$ are shown by the dashed lines. The crossing points of the solid and
dashed 
lines are the candidate locations for $\beta$ Cephei on the HR-diagram.
}
\label{fig:2}
\end{figure}

By sweeping out evolutionary models of various masses, we search for stellar
models whose radial and quadrupole mode eigenfrequencies coincide with the
observed frequency $f_1$ and $f_2$. Figure \ref{fig:2} shows the series of
these models on the HR diagram, calculated by using Pacy\'nski's (1970) program
($X_0=0.73$ and $Z_0=0.02$). 
The two dashed lines show the models of which
the quadrupole mode has the same frequency as $f_2$, and the solid
lines show the models whose radial 
mode frequency
coincides with $f_1$. $\beta$ Cephei must be on one of the crossing points of
the solid lines and the dashed lines.  The candidate is (i) a $\sim 18 M_\odot$
star at the middle of the hydrogen core-burning phase, or (ii) a $\sim 12
M_\odot$ star near the turning point, or (iii) a $\sim 9 M_\odot$ star at the
late stage of the hydrogen core-burning phase.

\section{Deduction of $i$ and $\beta$}
\label{ibeta}
The oscillation mode $f_1$ is identified as the radial fundamental mode in the
case (i) or (iii) in the previous section. On the other hand, it is identified
as the radial first harmonic in the case (ii). First, let us consider the case
(iii). The radius of the star for the case (iii) is estimated from the stellar
evolution calculation as $R \simeq 6.5 R_\odot$. 
Since in our consideration $2\pi/\Omega = 6$ days
and $V_{\rm e}\sin i \simeq 25$ km/s, this means 
$i\simeq 30^\circ$ or $150^\circ$.
Combining this with the first estimate of $|\tan\beta \tan i| \simeq 2.2$
deduced from the power spectrum, we obtain $\beta \simeq 75^\circ$ or
$105^\circ$.  The estimate $\tan\beta
\tan i \simeq -6.14$ deduced from the recent magnetic field measurements 
results in $\beta \simeq 95^\circ$.  We obtain that the solutions
from both estimates of $\tan\beta \tan i$ point towards an almost equal
geometry. The orientation of the magnetic axis can only be derived from the
measurements of the magnetic field. We conclude that the angle between the
rotation axis and the magnetic axis amounts to some $100^\circ$ in $\beta$
Cephei in the case of scenario (iii).

If we take the case (ii), the radius of the star is larger than the case (iii)
($R\simeq 8.5 R_\odot$),
and hence $i$ becomes smaller ($i\simeq 20^{\circ}$ or $i\simeq 160^{\circ}$) 
and $\beta$ becomes close to $90^\circ$ ($\beta\simeq 80^{\circ}$ or
$100^{\circ}$ if $|\tan\beta\tan i|\simeq 2.2$ and $\beta\simeq 93^\circ$ if
$\tan\beta\tan i \simeq -6.1$).
Though the case (ii) cannot be ruled out, we think the case (iii) is more
likely because the radial fundamental mode is more easily excited. The case (i)
seems unlikely because of a high mass required.

In the case of $m\neq 0$, the frequency of the quadrupole mode is shifted by
the Coriolis force by $mC\Omega\cos\beta$ (Shibahashi \& Takata 1993).  Here
$C$ is determined by the equilibrium structure and the eigenfunction, and it is
of the order of $\sim 0.1$ for the low order p-modes of $\ell=2$.  Since
$|\cos\beta| \sim 0.17$, the frequency shift due to the Coriolis force is so
small that we do not need to change the conclusion about the evolutionary stage
discussed in the previous section.

\begin{figure}[hbt]
\epsfxsize=8cm
\centerline{\epsfbox[76 210 792 792]{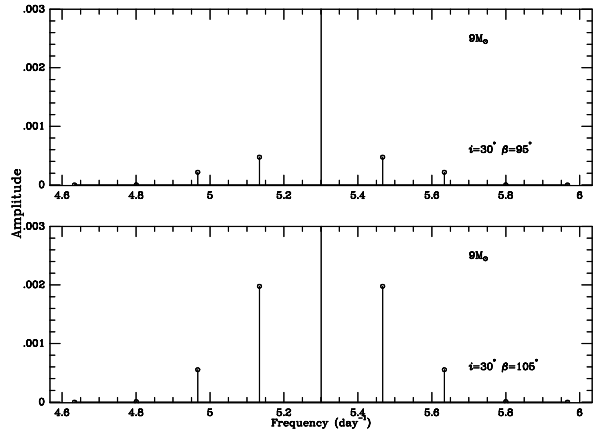}}
\caption{\footnotesize
Theoretically expected power spectrum. The model is a $9 M_\odot$ star at the
late stage of the hydrogen core-burning phase. The mode $f_1$ is identified as
the radial fundamental mode. The angle between the rotation axis and the
line of sight is assumed to be $30^\circ$ and the magnetic axis is assumed to
be inclined to the rotation axis by $95^\circ$ (upper panel) and $105^\circ$
(lower panel). The magnetic field strength is assumed to be 
$B_{\rm obs, max} = 90$ G, and the field is assumed to be 
a dipole + quadrupole field,
and the latter contribution is 10\%.
}
\label{fig:3}
\end{figure}

We have adopted the case (iii), and have calculated the theoretically expected
power spectra and compared them with the observations.  Figure \ref{fig:3} 
was calculated with an assumption of $B_{\rm obs, max} = 90$ G,
$i = 30^\circ$, and $\beta=95^\circ$ (the upper panel) and $\beta = 105^\circ$
(the lower panel). The magnetic field was assumed to be mainly dipolar with a
10\% contribution from a quadrupole component. (Note that the
combination of $i$ and $\beta$ is reversible.) Figure
\ref{fig:3} resembles the observed power 
spectrum, and it implies 
that our identification of the pulsation modes, 
the evolutionary stage of the star, and the geometrical configuration are
reasonable.

\section{Discussion}

The UV line variation of $\beta$ Cephei implies that the rotation period is
either 6 days or 12 days with the latter being more likely as the UV line
equivalent-width reveals a variation with alternating deep and less-deep minima
(Henrichs et al. 1993). Based on this observation, Henrichs et al.\ (1999)
prefer an oblique rotator model with a dipolar magnetic field and
with a rotation period of 12 days, the more so since this is the main period
found in their 14 recent observations of the averaged value of the magnetic
field over the stellar disk.  Such a model can explain the UV data if an
equator-on view is assumed.  However, if the rotation period is 12 days and the
magnetic field is a pure dipole, then the power spectrum of the line-profile
variations must show a quintuplet of an equal spacing of 1/12 day$^{-1}$
for $i\neq 90^\circ$ and $\beta\neq 90^\circ$, or a triplet of an equal spacing
of 1/6 day$^{-1}$ for $i = 90^\circ$ or $\beta = 90^\circ$; a
quintuplet fine structure of an equal spacing of 1/6 day$^{-1}$ is in that case
unrealistic.

\begin{figure}[ht]
\epsfxsize=8cm
\centerline{\epsfbox[76 210 792 792]{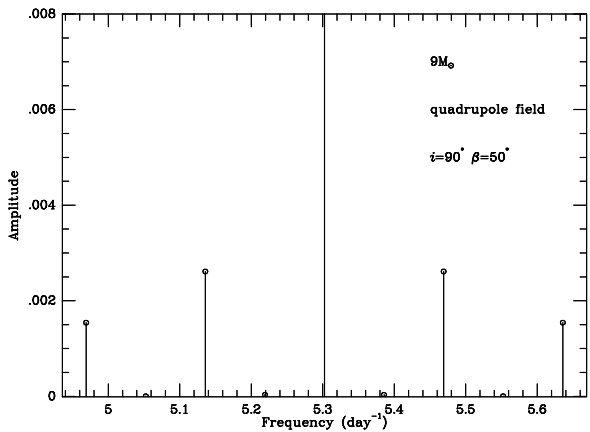}}
\caption{\footnotesize 
The same as figure 3 but for the almost quadrupole magnetic field and 
$i = 90^{\circ}$ and $\beta = 50^{\circ}$. The value of $i$ has been chosen 
so that it is consistent with $R \simeq 6.5 R_\odot$, 
$2\pi/\Omega$ = 12 days and $V_{\rm e}\sin i = 25$ km/s.
The value of $\beta$ has been chosen so that $r \simeq -1.4$ with an
assumption of a quadrupole magnetic field and $i = 90^{\circ}$.
The dipole field contribution is assumed to be only 0.01\%.
}
\label{fig:4}
\end{figure}

In order to get an apparent quintuplet of the spacing of 1/6 day$^{-1}$ with
the rotation period of 12 days from the oblique pulsator model, we have to
assume that the magnetic field is almost entirely quadrupolar rather than
dipolar and choose an
appropriate geometrical configuration to give only the five components among
the nonuplet
fine structure of the visible amplitude. 
The former condition is necessary, because, otherwise, 
the eigenfunction would have an $\ell=1$ component, which would induce 
a pair of peaks
separated from the central peak in the power spectrum by 1/12 day$^{-1}$.
In the case of a pure quadrupole
magnetic field, the eigenfunction is characterized by means of a superposition
of the spherical harmonic with $\ell=m=0$ and those of $\ell=2$ and $m=0$
and $\ell=4$ and $m=0$ (Takata \& Shibahashi 1994).
The $\ell=4$ component induces a nonuplet
fine structure.  But, in the case of $i \simeq 90^{\circ}$ or 
$\beta \simeq 90^{\circ}$, both of the
amplitudes at $\omega\pm \Omega$ and at $\omega\pm 3\Omega$ happen to 
become much smaller
than those at $\omega\pm 2\Omega$ and $\omega\pm 4\Omega$, and the fine
structure looks like a quintuplet with an equal spacing of 1/6 day$^{-1}$
(see figure \ref{fig:4}).
The combination of $2\pi/\Omega =$ 12 days and $V_{\rm e}\sin i \simeq 25$
km/s indeed leads to $i \simeq 90^{\circ}$, and one might consider that 
this would be favorable to explain the apparent quintuplet fine structure.
However, in the case of a pure quadrupole
magnetic field, the observed magnetic field strength should
vary as 
\begin{equation}
	B_{\rm obs} \propto P_2(\cos\Theta) \propto 3\cos^2 \Theta -1, 
\end{equation}
where 
\begin{equation}
	\cos\Theta = \cos\beta\cos i - \sin\beta\sin i \cos \Omega t.
\end{equation}
Then, in the case of $i \simeq 90^{\circ}$, the observed magnetic field 
strength is expected to vary with a period of 6 days rather than 12 days, and 
this is in contradiction with the observation (Henrichs et al. 1999).

In order to solve the controversy about the rotation period of the star, it is
highly desirable that numerous new magnetic field measurements be performed
over a much longer time base than achieved so far. We note that the older
magnetic field measurements pointed towards very different values of the mean
field, ranging from 70\,G up to 800\,G (Rudy \& Kemp 1978, Veen 1993).  The new
data obtained by Henrichs et al. (1999), however, are of much
better quality. It would be extremely important to confirm the results by
Henrichs et al. (1999) and to achieve a better precision of the
strength and the geometry of the magnetic field.  This would allow a critical
evaluation of the current theory of pulsations in hot magnetic stars.

\acknowledgments
We would like to express our sincere thanks to Dr.\ John Telting 
for helpful discussions.  This work was supported in part by a
Grant-in-Aid for Scientific Research of the Japan Society for the Promotion of
Science (No. 11440061).

\end{document}